\documentclass[a4paper]{article}
\usepackage[utf8]{inputenc}
\usepackage{hyperref}
\usepackage{csquotes}
\usepackage{indentfirst}
\usepackage{makecell}
\usepackage{graphicx}
\graphicspath{ {images/} }
\MakeOuterQuote{"}

\title{Mediation Analysis in Online Experiments at Booking.com: Disentangling Direct and Indirect Effects}
\author{Bahattin Tolga Öztan \\ tolga.oztan@booking.com
	\and Zoé van Havre \\ zoe.vanhavre@booking.com
	\and Caio Gomes \\ caio.gomes@booking.com
	\and Lukas Vermeer \\ lukas.vermeer@booking.com}
\date{October 2018}

\begin{document}

\maketitle

\begin{abstract}

Online experimentation is at the core of Booking.com’s customer-centric product development. While randomised controlled trials are a powerful tool for estimating the overall effects of product changes on business metrics, they often fall short in explaining the mechanism of change. This becomes problematic when decision-making depends on being able to distinguish between the direct effect of a treatment on some outcome variable and its indirect effect via a mediator variable. In this paper, we demonstrate the need for mediation analyses in online experimentation, and use simulated data to show how these methods help identify and estimate direct causal effect. Failing to take into account all confounders can lead to biased estimates, so we include sensitivity analyses to help gauge the robustness of estimates to missing causal factors.

\end{abstract}

\section*{Introduction}

At Booking.com, one of the key ingredients to customer-centric product development is not (just) bright minds having great ideas, but collecting the evidence to support these ideas. We test each idea addressing a customer pain point via an AB test using our in-house experiment platform. This platform is able to test thousands of changes simultaneously, with real customers, collecting data on the outcome within minutes of being implemented\cite{Kaufman2017}. However, as Booking.com grows to more countries, more languages, and our products grow in scope from hotels to other types of accommodations, cruises, car rentals, and more, our product development procedures must also become more flexible and more sensitive to the interactions of these many goals.

A randomised controlled trial (AB test) helps us assess the causal effect of the implementation of an idea (from now on referred to as treatment) on some desired outcome. Even though we can never calculate the outcome for a given person under both exposed and not exposed conditions, we can still get an unbiased estimate of the effect of the treatment on the intended population, referred to as the Average Treatment Effect (ATE)\cite{Rubin1974}. If the ATE was all we cared about, we would be done. However, at Booking.com, we care about two things:
\renewcommand{\labelenumi}{\roman{enumi}}
\begin{enumerate}
	\item The ATE, \textit{the total effect} of our treatment on the outcome variable
	\item The \textit{mechanism of change}, how the treatment affected our visitors' behaviour and, as a result, how the outcome variable changed.
\end{enumerate}
We value ii over i. This is best illustrated by a working example.

\section*{Experiment: Reducing ‘Cancellations per visitor’}

Cancellations stemming from unclarity around accommodation policies, facilities or prices lead to a bad customer experience, which we want to avoid as a general principle. In addition, cancellations make our partners’ availability calculations more difficult and lead to a bad partner experience as well. A confused, dissatisfied customer or partner is more likely to call our customer service, which also increases our customer service agents’ load. 

\begin{figure}[h]
\centering
\includegraphics[width=\linewidth]{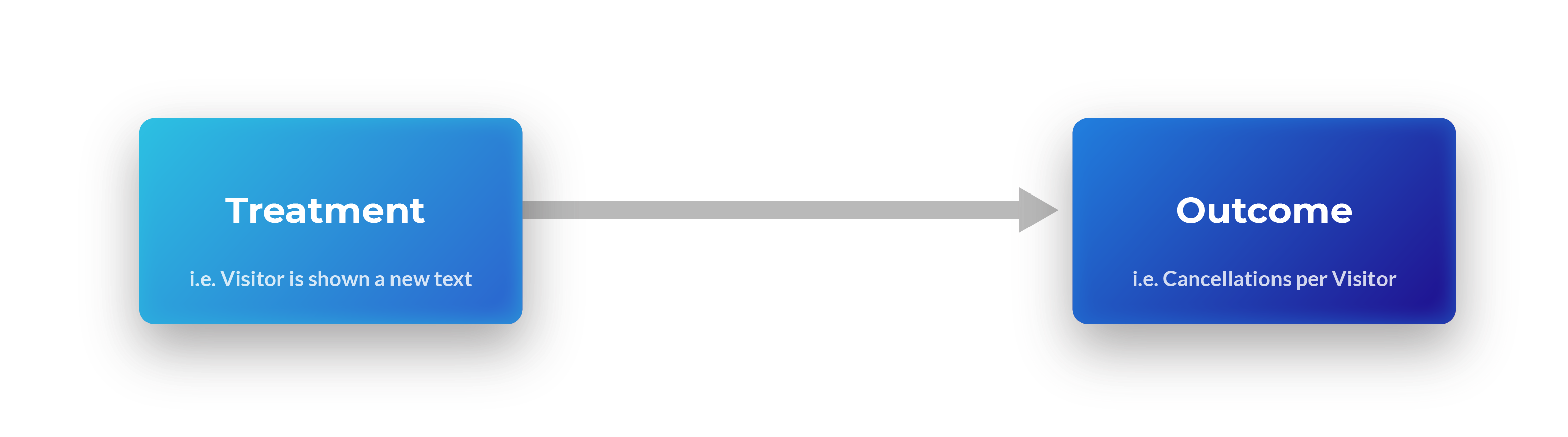}
\caption{A graphical representation of a treatment affecting an outcome.}
\label{fig:stage0}
 \end{figure}

We have teams working on solutions to address the pain points of customers and partners regarding cancellations. In one experiment, one such team might change the design of a page to bring more clarity to potential guests before they make a reservation, adding a text box containing an explanation of the property policies. Their goal is to make guests more aware of the prices around their trips, so that they have a lower chance of cancelling later, resulting in a reduction of the metric ‘cancellations per visitor’. We can represent such a scenario visually using the causal graph in Figure \ref{fig:stage0}.

If the only goal was to reduce cancellations, the experimenter could go ahead and use the ATE to see if this was achieved, testing the difference between the average cancellations per visitor between the control and treatment group. However, this would never tell us how this result was achieved. The learning comes with understanding the mechanism of change, and teams need this understanding to explore other ideas or abandon those that don't work.

\section*{Modelling the flow of effects}

\begin{figure}[h]
\centering
\includegraphics[width=\linewidth]{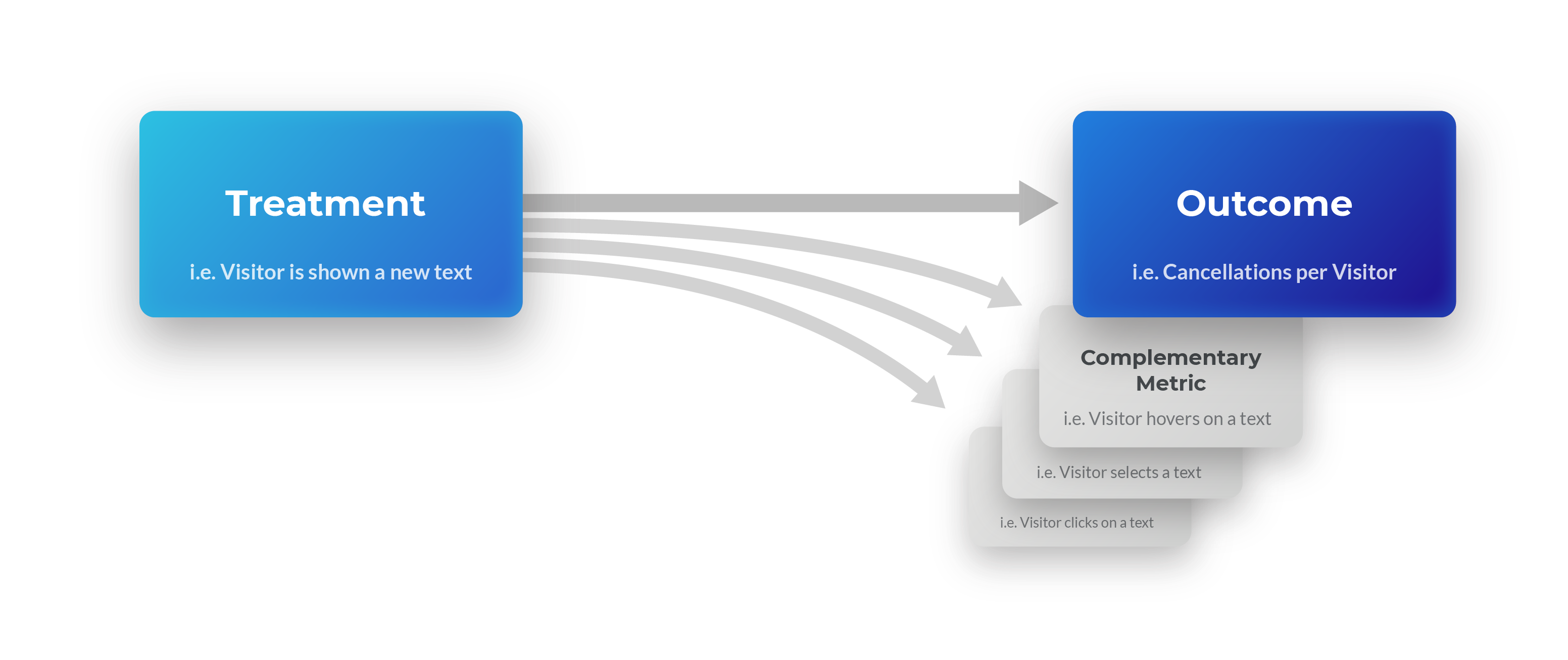}
\caption{A graphical representation of a treatment affecting an outcome and several complimentary metrics.}
\label{fig:stage1}
 \end{figure}

We encourage teams to monitor complimentary metrics that can provide additional support for their hypothesised mechanism. For the example above, as the new information is displayed in a text box, a supporting metric might be if visitors hovered on the text box or not. We can extend the causal graph from Figure \ref{fig:stage0} to include these supplementary metrics as shown in Figure \ref{fig:stage1}. 
 
Here the team is explicitly checking one specific mechanism which is the reduction of cancellations complemented  by a hover on the text box. If hovers increase and cancellations decrease, we understand better the mechanism. However, if cancellations change without the number of hovers changing then either an unforeseen mechanism is at work or we are dealing with a false positive. Either way, this additional insight can help the team better understand how their treatment is affecting visitor behaviour.

\section*{Direct \& Indirect Treatment Effects}

\begin{figure}[h]
\centering
\includegraphics[width=\linewidth]{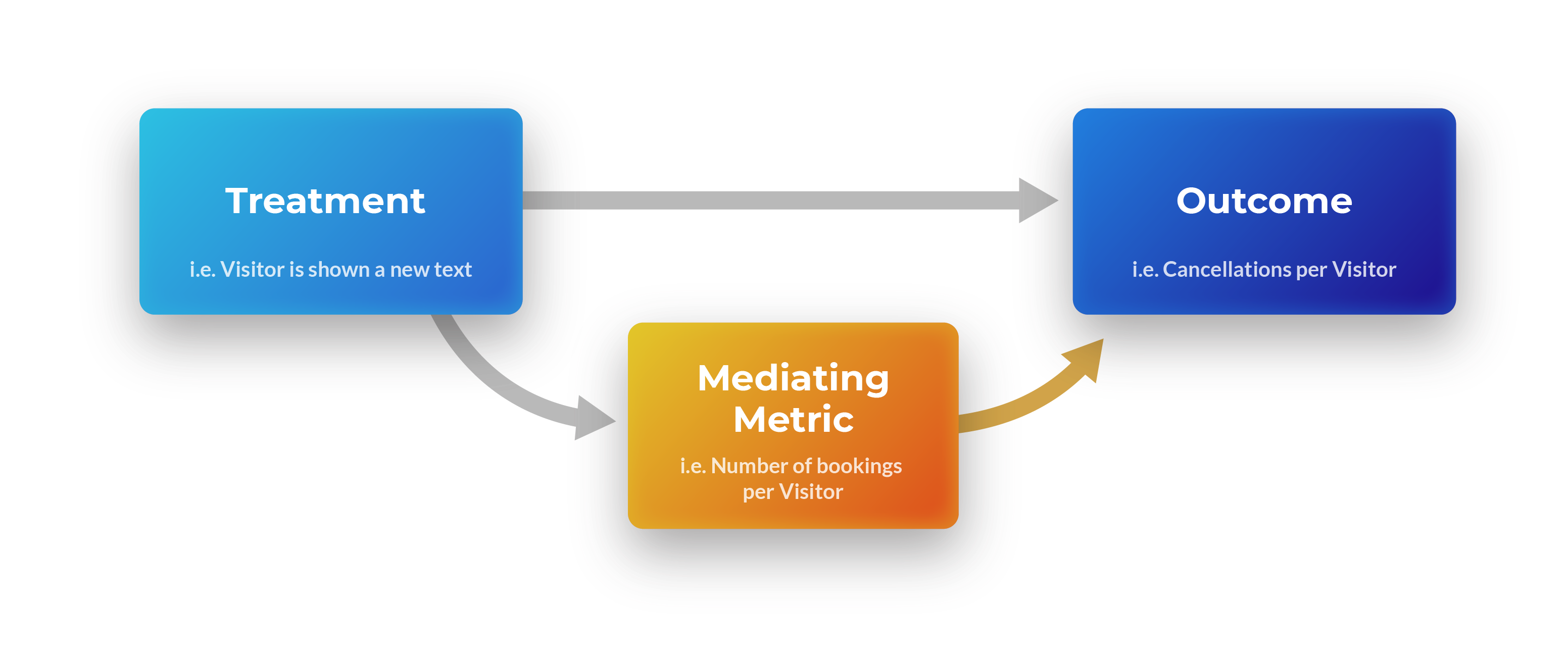}
\caption{A graphical representation of a treatment affecting an outcome directly, as well as through a mediating metric.}
\label{fig:stage2}
 \end{figure}

This approach gets more complicated when the metric that will help explain the mechanism is directly related to the outcome variable, while treatment is also expected to directly affect the outcome variable. For instance, if we observed a decrease in cancellations per visitor, but the number of bookings was also reduced. Did the reduction in cancellations originate from the new feature saving customers from making bookings, most of which would have been cancelled anyway, or did it inadvertently scare off previously satisfied customers from making bookings at all? Figure \ref{fig:stage2} expresses this mediation scenario graphically.

In this case, Bookings is a direct parent of the child metric Cancellations, meaning any change in bookings will automatically cause a change in cancellations. If you have more bookings, you have more opportunities for one of the bookings to get cancelled. If you have fewer bookings, you have fewer opportunities for a cancellation to happen. 

The Average Treatment Effect in this example can be broken down into two effects: 

$$ \textnormal{ATE} = \textnormal{Direct Effect} + \textnormal{Indirect Effect} $$

The Indirect Effect is the effect of treatment on cancellations via bookings and the Direct Effect is the effect of treatment on cancellations directly. A standard AB test will only tell us the sum of these two effects (the ATE) which is a very crucial quantity. If cancellations increase beyond an amount that the company can tolerate, they may not care if the increase is due to additional bookings or the treatment's direct effect. However, in most cases, to be able to understand the mechanism and to be able to make a decision, we need to disentangle these two effects.

\section*{Confounders}

\begin{figure}[h]
\centering
\includegraphics[width=\linewidth]{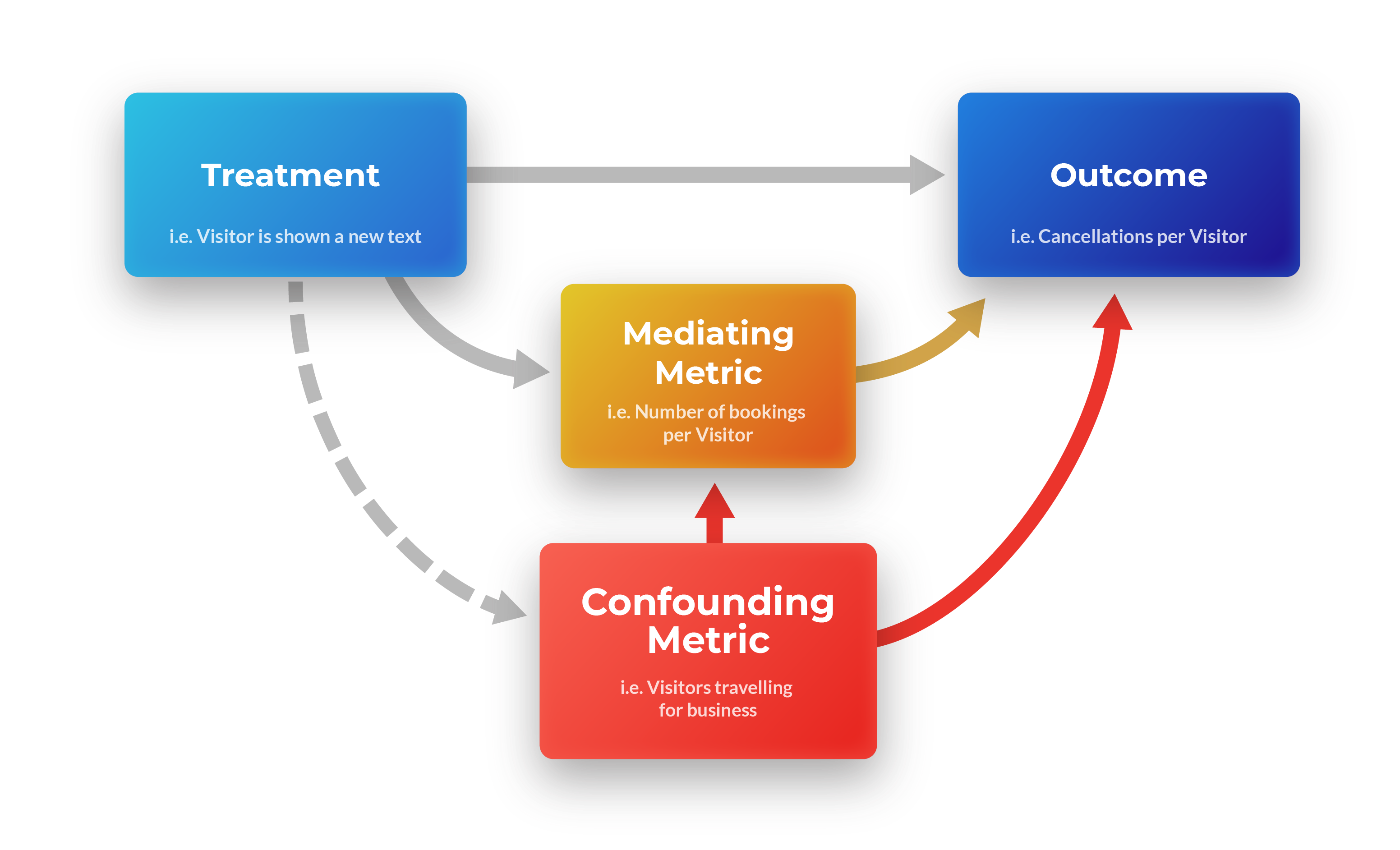}
\caption{A graphical representation of a treatment affecting an outcome directly, as well as through a mediating metric and a confounder.}
\label{fig:stage3}
 \end{figure}

Another lurking problem is that of known and unknown confounder variables. In the example above, imagine some visitors to the site are business travellers. We will use the causal graph with one confounder (‘is visitor travelling for business’) shown in Figure \ref{fig:stage3} as a running example in this analysis.
 
Confounders can add a layer of non-identifiability to the problem of interpreting the causal effect. While methods do exist which enable us to adjust for a post-treatment covariates, the moment we do this we lose the unbiasedness of the causal effect unless we can control for all potential confounders. The reason for this is that the confounders and treatment are not independent conditional on the mediator variable, bookings\cite{Pearl2009}.

By design, we randomise the visitors so that treatment and whether a visitor is a business traveller or not are independent. However, they are not conditionally independent. Imagine a scenario in which business visitors book more and also visitors in the treatment group book more. If we know how many bookings a visitor has then knowing if they are in treatment or not would give us information about their likelihood of being a business traveller or not, and vice versa.

\section*{Methods}

The method we employ follows Imai et al.\cite{Imai2011}. Given a set of pre-treatment confounders X, we assume the following, known as the sequential ignorability assumption:

$$ Y_{i}(t',m),M_{i}(t) \perp T_{i}|X_{i}=x $$
$$ Y_{i}(t',m) \perp M_{i}(t)|T_{i}=t,X_{i}=x $$
 
The first equation is guaranteed to hold by design as the context we work in is randomised online experiments. The second equation says the mediator variable is ignorable conditioned on the pre-treatment confounders X. Looking at it from the perspective of Directed Acyclic Graphs notation, we assume that all the back-door paths we may be opening by conditioning on the mediator variable will be blocked by the set of pre-treatment variables X. If this assumption holds, then the direct and indirect effects of treatment on the outcome variable are identified and we can estimate them non-parametrically\cite{Imai2011}.

For the estimation of causal effects, we use two-stage modeling as proposed in Imai et al.\cite{Imai2011,Tingley2014}. The form of these models are not important as long as sequential ignorability holds; so we use generalized linear models that are suitable for the outcome variables. 

\section*{Results}

We simulated 100,000 data points with one pre-treatment confounder, whether a visitor is a business traveller or not, where half the visitors are randomly assigned to treatment and half to control groups and within each group, the probability of a visitor being a business traveller is 0.4. On average a business traveller makes 1 booking while a non-business traveller on average makes also 1 booking, draws coming from Poisson distributions. The treatment adds 2 more bookings for business travelers but doesn’t affect at all bookings of non-business travellers. In addition, treatment has no direct effect on cancellations; therefore, cancellations per booking stays the same for each group; 14\% for business travellers and 7\% for non-business travellers, drawn from binomial distributions with said probabilities.

\begin{table}[htb]
\caption{Simulation parameters.}
\label{table:parameters}
\begin{tabular}{rrrrrr}
\thead[bc]{Treatment} & \thead[bc]{Business\\Traveler} & \thead[bc]{Share of\\Visitors} & \thead[bc]{Booking per\\Visitor} & \thead[bc]{Cancellations per\\Booking} & \thead[bc]{Cancellations per\\Visitor} \\
\hline
0 & 0 & 0.30 & 0.99 & 0.07 & 0.07 \\
0 & 1 & 0.20 & 0.99 & 0.14 & 0.14 \\
1 & 0 & 0.30 & 1.00 & 0.07 & 0.07 \\
1 & 1 & 0.20 & 3.00 & 0.14 & 0.42 \\
\end{tabular}
\end{table}

As seen in Table \ref{table:parameters}, the treatment has no direct effect on cancellations as the cancellation rates per booking stay the same for the subpopulations. All the treatment does is make \textit{business travellers} book more, from 1 booking per visitor on average to 3 on average. In this scenario, adding the number of bookings as a covariate in a regression model yields a positive regression coefficient for the effect of treatment on cancellations even though there is no direct effect of treatment on cancellations. Since there is a pre-treatment confounder and treatment and bookings are not conditionally independent, this approach gives a biased result. However, using the two-stage method and including the pre-treatment confounder in the models, we get direct effects close to zero, as expected. Table \ref{table:effects} contains point estimates for no adjustment, adjustment in a linear regression, and two-stage model results. Results in Table \ref{table:effects} further assume the experiment was run for 30 days in order to bring the point estimates down to familiar per day units.

\begin{table}[htb]
\caption{Direct effect of treatment on cancellations per day.}
\label{table:effects}
\centering
\begin{tabular}{lrr}
\thead[bl]{Method} & \thead[bl]{Effect} & \thead[bl]{p-value} \\
\hline
No Adjustment - ATE & 384 & 0.00 \\
Linear Regression & 65 & 0.00 \\
\makecell[tl]{Effect on Business Bookers\\2-Stage Method} & 6.143 & 0.32 \\
\makecell[tl]{Effect on Non-Business Bookers\\2-Stage Method} & 1.388 & 0.68 \\
\end{tabular}
\end{table}

Next, we do a sensitivity analysis to see how robust the point estimates are to missing covariates. Data can never tell us whether we have successfully taken into account all pre-treatment confounders. However, a sensitivity analysis can tell us how robust the estimates are. In our simulation data, we know there is only one pre-treatment confounder. So we expect the estimate to be quite robust when we include this variable in our models, and not robust when we omit it. 

\begin{figure}[h]
\centering
\includegraphics{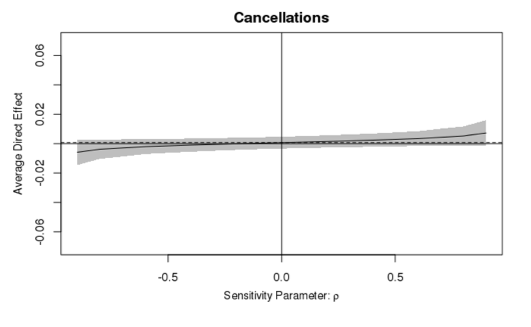}
\caption{Sensitivity plot for direct effect including confounder.}
\label{fig:with}
\end{figure}
 
\begin{figure}[h]
\centering
\includegraphics{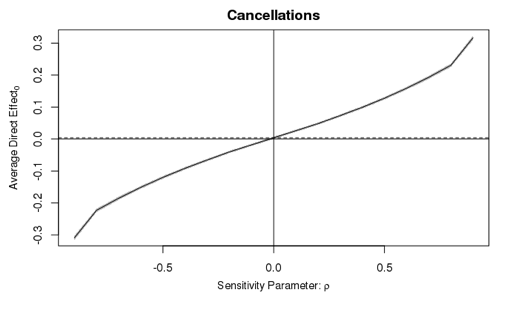}
\caption{Sensitivity plot for direct effect omitting confounder.}
\label{fig:without}
 \end{figure}

Figure \ref{fig:with} shows how the direct effect of the treatment on cancellations changes with the correlation of the error terms of stage 1 and stage 2 models when we include the pre-treatment covariate. Observe that even when the correlation term $\rho$ on the x axis is at the extremes, -0.9 and 0.9, the 95 percent intervals still include 0, meaning we can never conclusively conclude that the effect is non-zero. On the other hand, when we omit the pre-treatment covariate, \textit{business booker}, the conclusions change drastically depending on the value of $\rho$; as shown in Figure \ref{fig:without}.

\section*{Discussion}

AB testing allows for rapid customer-centric product development, but is likely to produce biased direct effect estimates, as the method cannot account for confounders, nor identify missing confounders. This leads to decisions being governed by hidden factors, which will at best inject randomness into the decision making process (costing time and effort), and at worst erode the quality of a product. For example, in the hypothetical scenario above, naive use of standard AB testing might lead to a product better tailored for business travellers, and by consequence shrink our customer pool significantly. If identifying the \textit{mechanism of change} is important for decision making, then we suggest (a) using multiple experiments to replicate findings and protect against false positives, (b) measuring important health metrics, as well as metrics known to be related (causally) to the outcome (and to include these in the model as demonstrated), and (c) performing a sensitivity analysis to check for missing confounders. We encourage experimenters to always interpret results in context; making decisions which keep in mind the big picture, the mechanism of change, and the long term impact on customers.

\section*{Acknowledgements}

The ideas put forward in this paper were greatly influenced by our work on the in-house experimentation platform at Booking.com, as well as conversations with colleagues and other online experimentation practitioners. In particular, this work would not have been possible if not  for the relentless challenging by Raphael Lopez Kaufman. Causal graph figures were crafted by our amazing designer Sergey Alimskiy.

\end{document}